%% file: paper.tex
\documentstyle[preprint,pre,aps,amssymb,epsfig]{revtex}
\tightenlines

\begin{document}

\draft
\title{Morphological phase transitions of thin fluid films on
chemically structured substrates}
\author{C. Bauer$^1$, S. Dietrich$^1$, and A. O. Parry$^2$}
\address{$^1$Fachbereich Physik, Bergische Universit\"at Wuppertal,
D-42097 Wuppertal, Germany}
\address{$^2$Department of Mathematics, Imperial College,
180 Queen's Gate, London SW7 2BZ, United Kingdom}
\date{1. April 1999}

\maketitle

\input{abstract}


\input{body}

%% file: abstract.tex
\begin{abstract}
Using an interface displacement model derived from a microscopic
density functional theory we investigate thin liquidlike wetting
layers adsorbed on flat substrates with an embedded chemical
heterogeneity forming a stripe. For a wide range of effective
interface potentials we find first-order phase transitions as well as
continuous changes between lateral interfacial configurations bound to and
repelled from the stripe area. We determine phase diagrams and discuss
the conditions under which these morphological changes arise.
\end{abstract}

%% file: body.tex
A variety of experimental techniques have emerged which allow one to
endow solid surfaces with a rich, well-defined and permanent chemical
pattern while keeping the surface flat on a molecular scale (see,
\emph{e.g.}, ref.~\cite{microstructuring}). An important application 
of these structures is
microfluidics~\cite{knightetal,grunze}, \emph{i.e.}, the guidance of tiny
amounts of adsorbed liquids along these chemical microstructures. This
enables one to control the microscopic flow of liquids on 
designated chemical channels and to faciliate the fabrication of
``chemical chips'' which may act as microlaboratories for the
investigation and processing of rare and valuable
liquids~\cite{service}. Although these applications involve 
dynamical processes, as a prerequisite it is important to investigate
these systems in thermal equilibrium as a function of the
thermodynamic parameters pressure (or, equivalently, chemical
potential $\mu$) and temperature $T$.
Some recent studies were concerned with the static behaviour of liquid
channels on chemical lanes within the micrometre range (see
refs.~\cite{lenzlipowsky} and \cite{gauetal}). On this scale the
morphology of the adsorbed liquid is determined by gross features such as the
various surface tensions involved. However, with the rapidly
proceeding miniaturisation of microstructures in mind, here we are
interested in a much smaller scale within
which details of the molecular forces become relevant \cite{sdsicily}.

\begin{figure}
\begin{center}
\epsfig{file=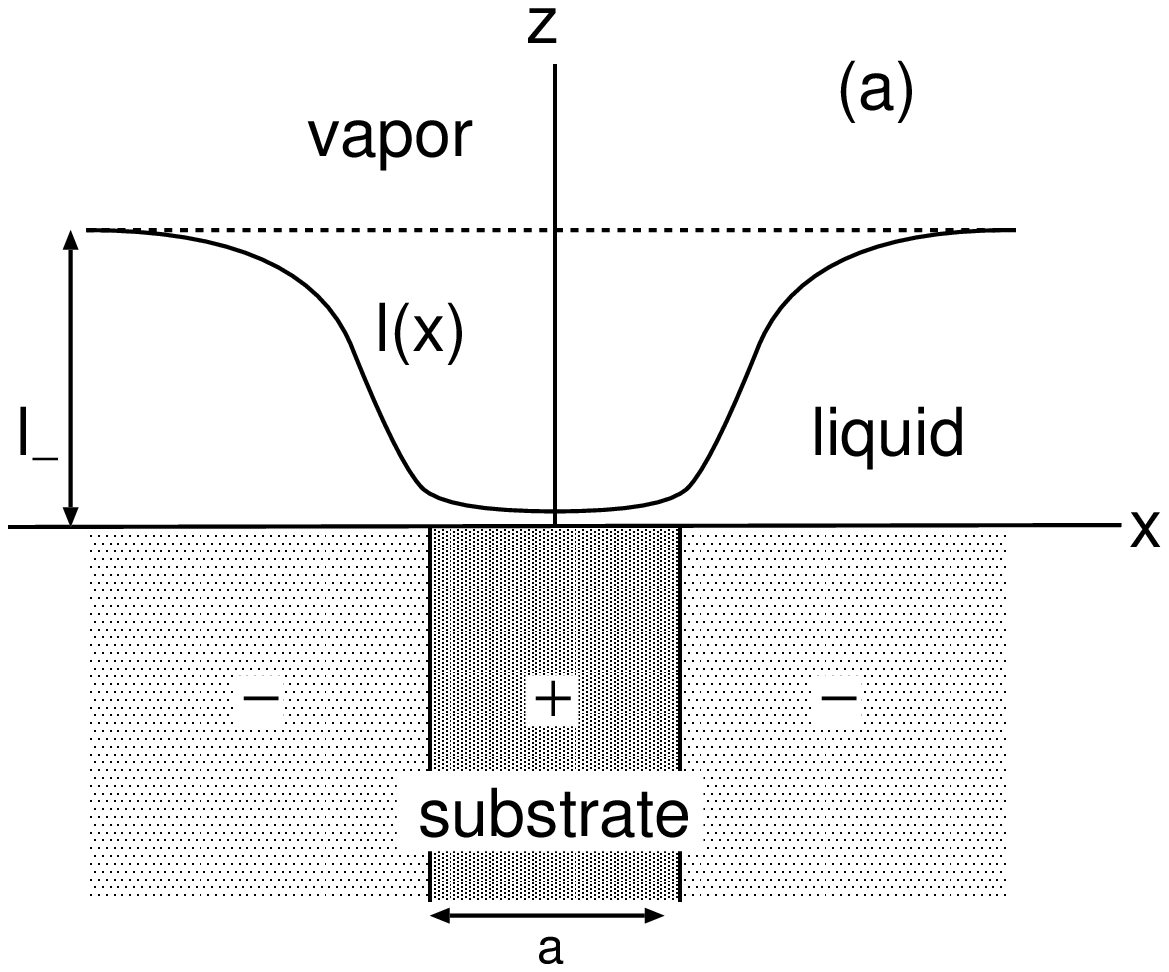, width=6.8cm}
\epsfig{file=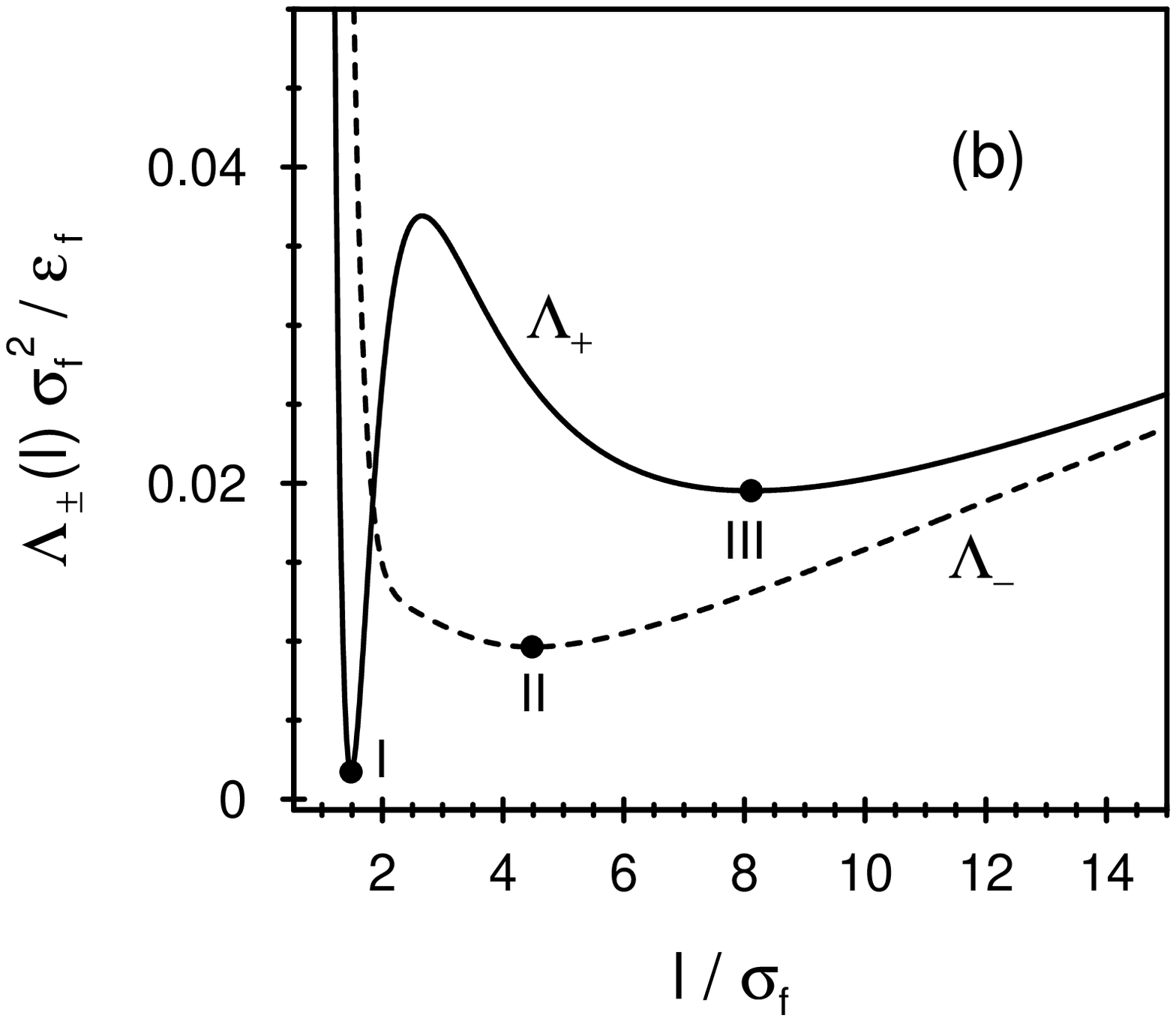, width=6.8cm, bbllx=25, bblly=380,
  bburx=495, bbury=795} 
\end{center}
\vskip 0.5cm
\caption{\label{f:system}
(a) Cross section of the morphology of a liquidlike wetting film covering a planar
substrate that contains a slab of different material. The substrate surface is
located at $z=0$ and the system is translationally invariant in the
$y$ direction. Viewed from the top the stripe region is filled with chemical
species denoted by ``$+$'' and extends from $x=-a/2$ to $x=a/2$, \emph{i.e.}, the
stripe width is $a$. $l_- = l(|x|\to\infty)$ is the equilibrium thickness of the
liquidlike wetting film corresponding to a flat, homogeneous substrate
composed of ``$-$'' particles alone. The stripe and the outer region favour a
thin and a thick wetting layer, respectively. (b) The effective interface
potentials $\Lambda_+(l)$ (full line) and 
$\Lambda_-(l)$ (dashed line) of the flat and \emph{homogeneous} substrates
composed of the chemical species ``$+$'' and ``$-$'',
respectively, are minimized for $l=l_+$ and $l=l_-$,
respectively. These interface potentials correspond to 
$k_BT/\epsilon_f=1.1$ and 
$\Delta\mu/\epsilon_f=0.003$ (as, c.f., in fig.~\ref{f:transition}). For
$\Delta\mu\to0$ the positions of the minima II and III are shifted towards 
larger values of $l$ and to $l=\infty$, respectively, whereas I remains
basically unchanged. The 
global minima I and II of $\Lambda_{\pm}(l)$ correspond to 
the equilibrium film thickness of the liquidlike film adsorbed on
the respective \emph{homogeneous}
substrate. $\Lambda_{\pm}(l\to\infty)$ increases linearly as 
$\Delta\mu\Delta\rho\,l$. $\Lambda_{\pm}(\Delta\mu=0,l=\infty)=0$ and
$\Lambda_{\pm}(\Delta\mu=0,l=l_{I,II})<0$, \emph{i.e.}, at
two-phase coexistence $\Delta\mu=0$ and at the above temperature both
types of substrates are only partially wet.}  
\end{figure}

The paradigmatic system considered here is a chemically heterogeneous surface
which, in top view, exhibits a single stripe. The substrate is flat and
composed of two different chemical species such that one of them
(denoted by ``$+$'') forms a single slab of width $a$ 
embedded in the other one (denoted by ``$-$''; see
fig.~\ref{f:system}(a)). Based on a density functional approach
\cite{bauer} it turns out that the 
liquid-vapour interface of a thin liquidlike layer in contact with the
wall interacts with the wall via an \emph{effective} interface potential
$\Lambda(x,l(x))$, where $l(x)$ is the local thickness of the layer at the
lateral position $x$. The substrate potential entering into
$\Lambda(x,l(x))$ has been obtained from a
pairwise summation over all substrate-fluid particle interactions
assuming sharp chemical steps at $x=\pm a/2$. Since the system we consider
here is translationally invariant in the $y$ direction, $\Lambda$ does
not depend on $y$. We take both the substrate-fluid ($s$) and the
fluid-fluid ($f$) interaction potential to be of Lennard-Jones type:
$\phi_{s,f}(r)=4\epsilon_{s,f}((\sigma_{s,f}/r)^{12}-(\sigma_{s,f}/r)^6)$. We
choose the two chemical species such that a flat, semi-infinite, and
\emph{homogeneous} substrate composed of each of the species \emph{alone} 
exhibits an effective interface potential $\Lambda_{\pm}(l)$ as depicted in
fig.~\ref{f:system}(b). (One may interpret these interface potentials as
corresponding to a ``hydrophilic'' ($\Lambda_-(l)$) and a ``hydrophobic''
($\Lambda_+(l)$) surface even for a nonvolatile liquid.)
For our choice of potential parameters the outer part of the substrate
undergoes a critical wetting transition at
$k_BT_w/\epsilon_f=1.2$, whereas a homogeneous
substrate filled with species corresponding to the stripe part exhibits a
first-order wetting transition at $k_BT_w/\epsilon_f\approx1.102$. For
the temperature $k_BT/\epsilon_f=1.1$ considered throughout the paper
both substrate types are only partially wet at bulk liquid-vapour 
coexistence $\mu=\mu_0$, \emph{i.e.}, $\Delta\mu=\mu_0-\mu=0$.

Within a simple interface displacement model, which can be derived
systematically from density functional theory (see ref.~\cite{bauer})
the equilibrium contour $l(x)$ of the liquid-vapour interface minimizes
the functional 
\begin{equation}\label{e:functional}
\Omega_s[l(x)] = \int_A dx\,dy\,\Lambda(x,l(x))
+ \sigma_{lg}\int_A dx\,dy\,\sqrt{1+\left(\frac{dl(x)}{dx}\right)^2}
\end{equation}
with the surface area $A=L_xL_y$ of the substrate surface and
$\Lambda(x,l)=\Delta\mu\Delta\rho\,l+\omega(x,l)$ where, as it turns out,
$\omega(x,l)=\sum_{i\geq2}a_i(x)l^{-i}$ for Lennard-Jones
potentials, $\Delta\rho=\rho_l-\rho_g$ is the difference in number
densities between the bulk phases, and $\sigma_{lg}$ is the surface tension
associated with the area of the liquid-vapour interface. Instead of the chemical
potential one may use the pressure $p$ of the bulk vapour phase as a
thermodynamic control parameter. In this case $\Delta\mu=0$
corresponds to $p=p_{sat}(T)$ at
which the vapour phase is saturated. In a pressure-temperature ensemble
one has to replace $\Delta\mu\Delta\rho$ by the pressure difference $\Delta p =
p_{sat}-p$. In eq.~(\ref{e:functional}) we have omitted the 
free energy contributions from the wall-liquid interface at $z=0$ which are
constant with respect to $l(x)$. Subtraction of the surface free 
energy $\Omega_s(l_-)$ corresponding to the homogeneous outer
(``$-$'') substrate yields the line contribution $L_y\Omega_l[l(x)] = 
\Omega_s[l(x)]-\Omega_s(l_-)$ to the free energy of the fluid
configuration associated with the presence of the
chemical stripe. $\Omega_s[l(x)]$ (or equivalently, $\Omega_l[l(x)]$) is minimized
numerically with respect to $l(x)$ yielding the equilibrium contour
within mean field theory.

\begin{figure}
\begin{center}
\epsfig{file=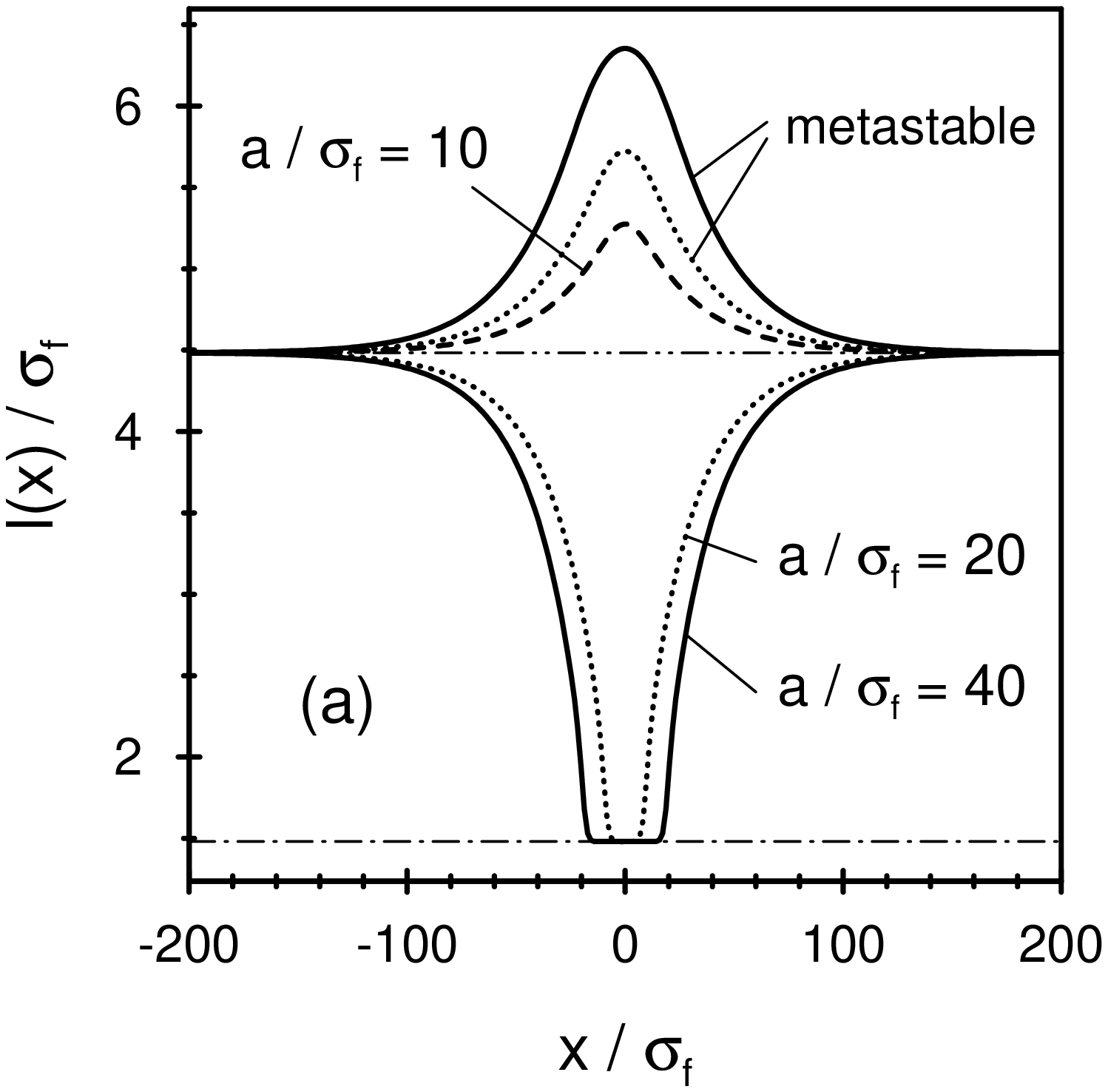, height=6.7cm, bbllx=90, bblly=335,
  bburx=530, bbury=775} 
\epsfig{file=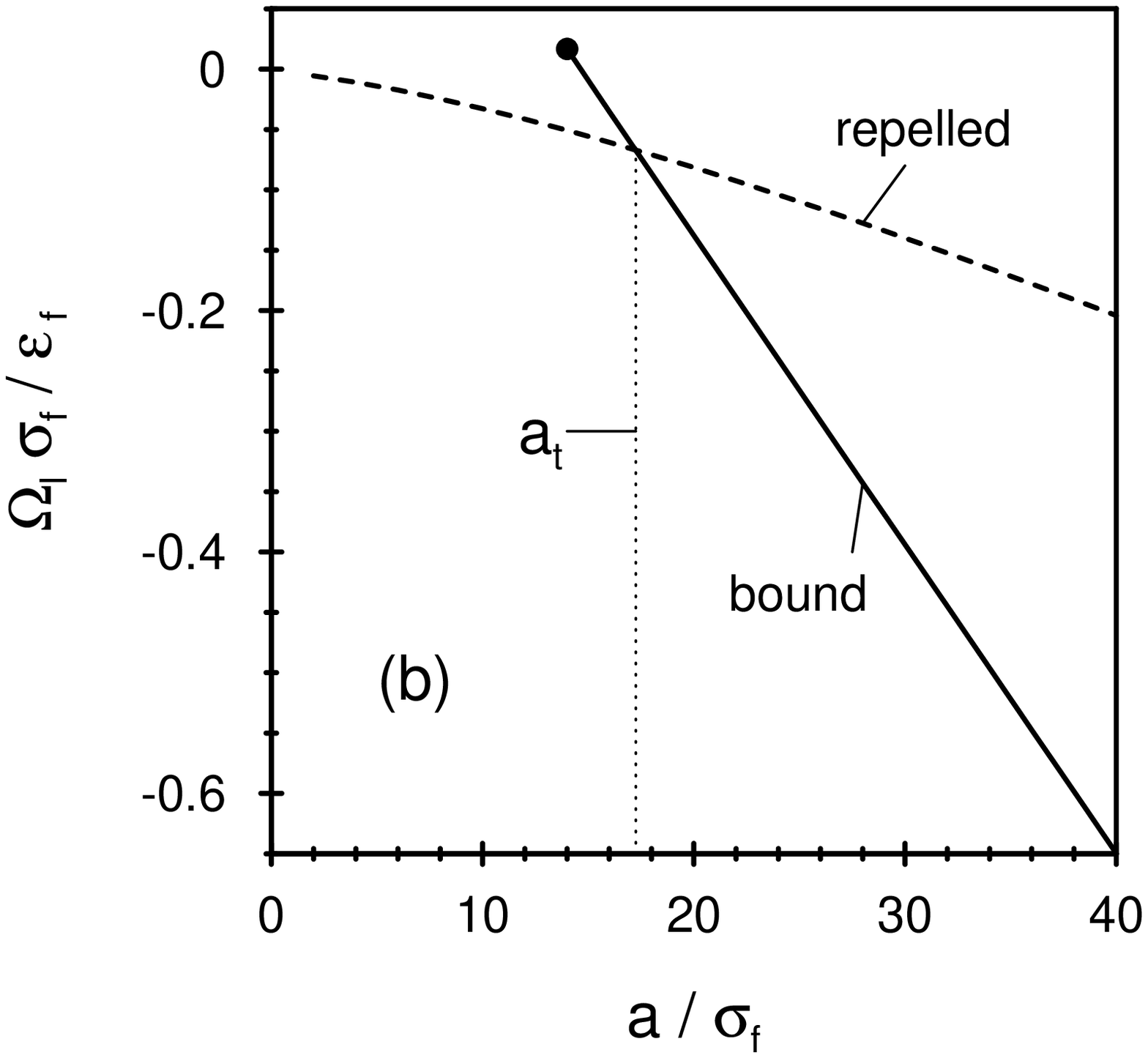, height=6.7cm, bbllx=55,
  bblly=335, bburx=520, bbury=775} 
\end{center}
\caption{\label{f:transition}
(a) Liquid-vapour interfaces on the stripe-shaped heterogeneity
calculated for $k_BT/\epsilon_f = 1.1$ and
$\Delta\mu/\epsilon_f=3\cdot10^{-3}$ based on the effective interface
potentials shown in fig.~\ref{f:system}(b). The interface
profiles are calculated for $a/\sigma_f=10$ (dashed line), $20$ (dotted
lines), and $40$ (full lines). For $a=10\sigma_f$ there is only a repelled
solution. The dash-dotted lines indicate the equilibrium film
thicknesses $l_+$ ($-\,\cdot$) and $l_-$ ($-\cdot\cdot$). (b) 
That part of the line contribution $\Omega_l$ to the grand canonical potential 
which depends functionally on $l(x)$, for the same choice of parameters as in
(a). At $a=a_t\approx17.4\sigma_f$ there is a first-order phase transition
from the repelled (dashed line) to the bound solution (full line). The
dot indicates the end of the metastable branch of bound
solutions for $a<a_t$. The metastable branch of repelled solutions
extends far to large values of $a$.}
\end{figure}

Using the interface potential $\Lambda(x,l)$ we find equilibrium interfacial
morphologies, pertinent examples of which are shown in 
fig.~\ref{f:transition}(a). This figure depicts the interface profiles for 
varying stripe widths $a$ at a fixed value
$\Delta\mu=0.003\epsilon_f$. Within a wide range of values for $a$
there are two different minimal interfaces, one ``bound'' to and
the other ``repelled'' from the stripe area.

Figure~\ref{f:transition}(b) displays the values of the line free
energy density $\Omega_l$ corresponding to the interface profiles shown in
fig.~\ref{f:transition}(a). For large $a$ the solution bound to the 
stripe area has a lower line free energy $\Omega_l$ than the
solution repelled from the stripe. As $a$ is decreased the
latter solution is favoured in terms of the free energy. Thus at some value
$a=a_t$ there is a phase transition between both interfacial
configurations. Due to the break in slope of $\Omega_l(a)$ at $a=a_t$
this transition is first
order. We refer to this phenomenon as ``morphological phase
transition'' in the sense that the interface profile $l(x)$ undergoes
an abrupt structural change.  

The repelled solution exhibits a coverage which is even larger than that
corresponding to the homogeneous ``$-$'' substrate. This
counterintuitive result shows that compared with a homogeneous
substrate, one can \emph{increase} the total
adsorption by the immersion of a slab of a material that favours \emph{thinner}
liquidlike films. The occurrence of this phenomenon persists if the
depth of the slab is not macroscopicly large as in fig.~\ref{f:system}(a)
but only molecularly small corresponding to a different material
within an imprinted overlayer covering a homogeneous underlying
substrate. This example shows that gradual changes in the architecture
of chemically microstructured devices can lead to abrupt changes in
the morphology of adsorbed liquids. We emphasise that such phase
transitions are not only of theoretical interest. Their existence
illustrates the care required to avoid the unwanted filling of the
nonwet space between liquid channels in such devices.

For $|x|\to\infty$ the interface profile $l(x)$ asymptotically
approaches the equilibrium film thickness $l_-$ corresponding to the
outer part of the substrate. If the stripe width is sufficiently
large, in the middle of the stripe area $l(x)$ also approaches the
equilibrium film thickness $l_+$ corresponding to the stripe part. In
this case, as expected, $l(x)$ smoothly interpolates between the two minima I and II
of the interface potentials of the respective homogeneous and flat
``$+$'' and ``$-$'' substrate taking full advantage of the deep
minimum I (see fig.~\ref{f:system}(b)). If the stripe width shrinks
this gain in free energy decreases accordingly and, moreover, the
relative cost for this benefit in terms of the associated increased
area of the liquid-vapour interface increases. For $a<a_t$ the loss of
free energy by occupying the higher minimum III instead of I is
outweighed by the gain in free energy due to a reduced area of the
liquid-vapour interface, leading to the repelled solution because the
position of the local minimum III occurs at a larger value of $l$ than
for II. The physical nature of this transition is similar to the
``unbending'' transition on a corrugated wall as described in 
ref.~\cite{rasconparrysartori}. As explained in
ref.~\cite{rasconparry} similar effects such as ``out-of-phase
behaviour'' can also occur on nonplanar walls if there are competing
minima of the effective interface potential. 

\begin{figure}
\begin{center}
\epsfig{file=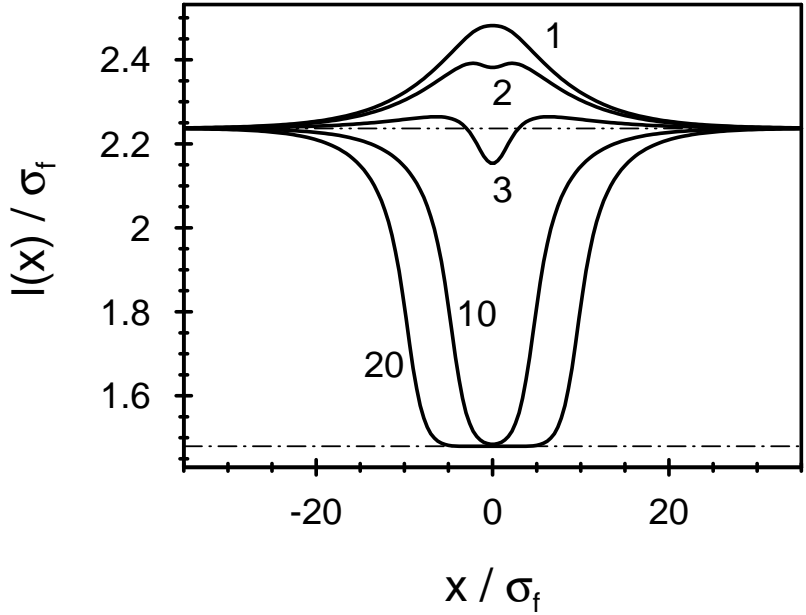, width=11cm, bbllx=60,
  bblly=330, bburx=515, bbury=685} 
\end{center}
\caption{\label{f:abovecritpoint}
Profiles of the liquid-vapour interface exposed to substrates as shown
in fig.~\ref{f:system}(a) for $k_BT/\epsilon_f=1.1$ and
$\Delta\mu=0.014\epsilon_f>\Delta\mu_c$. The 
numbers accompanying each graph indicate the corresponding value of
the stripe width $a$ in units of $\sigma_f$. The same set of
interaction parameters is used as for fig.~\ref{f:transition}. The
dash-dotted lines indicate the equilibrium film thicknesses $l_+$
($-\,\cdot$) and $l_-$ ($-\cdot\cdot$) (see
fig.~\ref{f:system}(b)). Since the system is above the critical point
$(\Delta\mu_c,a_c)$ of the morphological phase transition there is 
only one stable solution for every value of $a$. There is no
first-order phase transition and $\Omega_l(a)$ is a smooth function.}
\end{figure}

\begin{figure}
\begin{center}
\epsfig{file=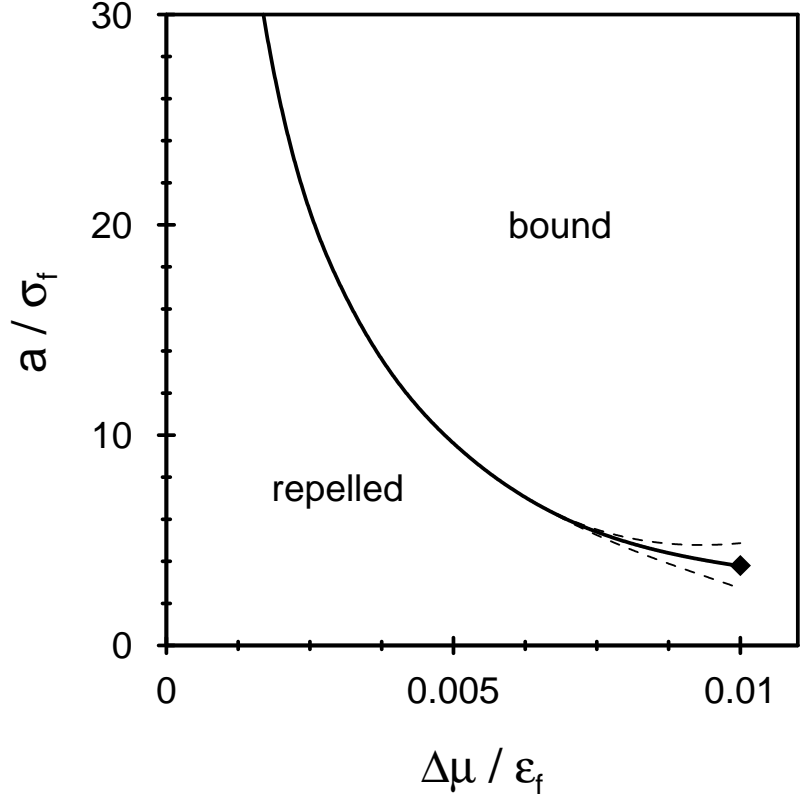, width=11cm, bbllx=75, bblly=325,
  bburx=520, bbury=775} 
\end{center}
\caption{\label{f:coexline}
Line of first-order phase transitions between solutions bound to
and repelled from the stripe area (full line). The same set of
interaction parameters is used as for figs.~\ref{f:transition} and
\ref{f:abovecritpoint}. Within mean field 
theory the line ends at a critical point which is indicated by the
diamond symbol. For decreasing $\Delta\mu$ the line of phase coexistence
extends further up to higher values of $a$. The dashed lines indicate
the region within which the rounding of the first-order transition due
to thermal fluctuations takes place. For decreasing values of $a$, \emph{i.e.},
upon approaching the mean field critical point, this rounding becomes more and
more important. Thus the fluctuations erase the phase transition near
$a_c$ and eliminate the associated critical point. For large values of
$a$ the fluctuation-induced rounding of the first-order phase
transition is practically not detectable.}   
\end{figure}

For values of $\Delta\mu$ which are larger than a certain critical
value $\Delta\mu_c$ the morphological phase transition does not occur,
\emph{i.e.}, there is only one stable solution for every value of
$a$. Figure~\ref{f:abovecritpoint} shows the behaviour of the
liquid-vapour interfaces for
$\Delta\mu=0.014\epsilon_f>\Delta\mu_{c}$. In this case the interface
profile gradually changes from a repelled 
configuration for small $a$ to a bound configuration for
large $a$. The corresponding line free energy $\Omega_l$ is an
analytic function that does not exhibit a cusp singularity and
associated metastable branches.
Figure~\ref{f:coexline} displays the line of first-order phase
transitions in the $(a,\Delta\mu)$ plane
at which the bound and the repelled solutions coexist. Within the present
mean field description this line ends at a critical
point with $\Delta\mu_c\approx 0.010\epsilon_f$ and $a_c \approx 3.8\sigma_f$. 

The number of degrees of freedom involved in the morphological phase
transition is proportional to the volume $L_y\,a\,(l_--l_+)$ and thus
quasi-onedimensional. Therefore the system cannot support a phase
transition in the strict sense of statistical mechanics. Actually the
fluctuations are so strong that at the phase boundary shown in
fig.~\ref{f:coexline} the system cannot sustain the sharp coexistence
between the bound and the repelled configuration. Instead the system
will break up into domains along the $y$ direction with alternating
regions of increased and reduced coverage with the positions of the
domain boundaries fluctuating. This leads to a rounding of the
first-order phase transition due to the aforementioned finite-size
effect in two directions and eliminates the critical point shown in
fig.~\ref{f:coexline}. However, for large values of $a$ or $l_--l_+$
the number of degrees of freedom involved becomes quasi-twodimensional
so that the rounding of the first-order phase transition sharpens up
turning it ultimately into a true discontinuity. Following the general
ideas of finite size scaling of first-order phase transitions
\cite{privmanfisher} one can estimate the width $2\delta\mu$ of the
rounding around the mean field location $\Delta\mu_t$ of the phase
transition according to the implicit equation
\begin{equation}
|\Delta\Omega_l(T,a,\Delta\mu_t(a)+\delta\mu(a))| \lesssim
 \frac{k_BT}{\xi_b}\exp\left(-\frac{a\Sigma_l(T,\Delta\mu_t(a))}{k_BT}\right)
\end{equation}
where $\xi_b$ is the bulk correlation length (which is of the order of
$\sigma_f$) and $\Sigma_l \approx \sigma_{lg}(l_--l_+)$ estimates the
cost in free energy to maintain an interface between two domains as
described above; $\Delta\Omega_l$ is the difference in line free
energies between the bound and the repelled solution (compare
refs.~\cite{gelfandlipowsky} and 
\cite{bieker}). A rough numerical estimate of $\delta\mu$ yields 
the boundaries of the smeared out transition region which are indicated by the
dashed lines in fig.~\ref{f:coexline}. Whereas far from the mean field
critical point the width of the transition region is exponentially small
such that the transition is quasi-first order, for $\Delta\mu$
increasing towards $\Delta\mu_c$ the rounding of the phase transition
becomes significant. 

In closing we note that the
only prerequisite for the occurrence of the morphological phase
transition described here is that the effective interface potentials
of the materials involved exhibit the gross features shown in
fig.~\ref{f:system}(b). This holds both for volatile and nonvolatile
liquids and does not depend on a particular type of interaction
potential or thermodynamic ensemble. Such morphological transitions
also occur for other shapes  
of chemical substrate heterogeneities such as periodic arrays of
stripes and circular or even irregular areas. A variety of
experimental techniques such as reflection interference contrast
microscopy (RICM) \cite{wiegandetal} or a force microscope used in
tapping mode \cite{herminghaus} are being developed which allow one to scan the
morphology of fluid films with a spatial resolution down to fractions
of a nanometre. The detection of the morphological phase transition
predicted here could serve as a promising testing ground for the
development of such experimental techniques.

\acknowledgements

C.B. and S.D. gratefully acknowledge financial support by the German
Science Foundation within the Special Research Initiative
\emph{Wetting and Structure Formation at Interfaces}.

\pagebreak

\pagebreak

\end{document}